\documentclass[fleqn,10pt]{wlscirep}
\usepackage[utf8]{inputenc}
\usepackage[T1]{fontenc}
\usepackage{float} 
\usepackage{multirow}
\usepackage{xr} 
\usepackage{graphicx}
\usepackage{graphicx} 
\usepackage{algorithm}
\usepackage{amsmath}
\usepackage{algpseudocode}
\usepackage{enumitem}
\usepackage{endnotes}

\title{Polarized Patterns of Language Toxicity and Sentiment of Debunking Posts on Social Media}

\author[1*]{Wentao Xu}
\author[1+]{Wenlu Fan}
\author[2+]{Shiqian Lu}
\author[3+]{Tenghao Li}
\author[4+]{Bin Wang}
\affil[1]{Department of Science and Technology of Communication}
\affil[2]{School of Computer Science and Technology}
\affil[3]{Department of Artificial Intelligence}
\affil[1,2,3,4]{University of Science and Technology of China, Hefei, 230000, China}
\affil[4]{Independent researcher, Beijing, 100000, China}

\affil[*]{corresponding author: myrainbowandsky@gmail.com}

\affil[+]{these authors contributed equally to this work}

\begin{abstract}

The rise of misinformation and fake news in online political discourse poses significant challenges to democratic processes and public engagement. While debunking efforts aim to counteract misinformation and foster fact-based dialogue, these discussions are often marred by language toxicity and emotional polarization. In this study, we examined over 86 million debunking tweets and more than 4 million Reddit debunking comments to investigate the relationship between language toxicity, pessimism, and social polarization in debunking efforts. Focusing on discussions of the 2016 and 2020 U.S. presidential elections and the QAnon conspiracy theory, our analysis reveals three key findings: (1) peripheral participants (1-degree users) play a disproportionate role in shaping toxic discourse, driven by lower community accountability and emotional expression; (2) platform mechanisms significantly influence polarization, with Twitter amplifying partisan differences and Reddit fostering higher overall toxicity due to its structured, community-driven interactions; and (3) a surprising negative correlation exists between language toxicity and pessimism, with increased interaction (e.g., replies) reducing toxicity, especially on Reddit. We further show that platform architecture affects the informational complexity of user interactions, with Twitter promoting concentrated, uniform discourse and Reddit encouraging diverse, complex communication. By incorporating an unprecedented scale of data from two distinct platforms, our findings highlight the importance of user engagement patterns, platform dynamics, and emotional expressions in shaping polarization in debunking discourse. This study offers actionable insights for policymakers and platform designers to mitigate harmful effects and promote healthier, more constructive online discussions, with broader implications for understanding the dynamics of misinformation, hate speech, and political polarization in digital environments.

\end{abstract}
\begin{document}

\flushbottom
\maketitle
%
%
\thispagestyle{empty}

\section*{Introduction}

In today’s digital age, the spread of fake news and misinformation has become a major concern, particularly in online political discourse \cite{pennycook2018fighting}. The proliferation of such misinformation has led to challenges in distinguishing fact from fiction, especially as social media platforms amplify polarized political discussions \cite{bakshy2015exposure}. These platforms, while providing access to diverse viewpoints, also contribute to echo chambers and ideological polarization. In this context, debunking false information is critical not only for restoring truth but also for promoting informed public engagement and protecting democratic processes \cite{lewandowsky2017beyond}. By systematically identifying and correcting misleading claims, debunking efforts can help foster more rational, fact-based political dialogue, counteracting the negative effects of misinformation and polarization \cite{friggeri2014rumor}.

As a defining characteristic of this polarized environment, language toxicity not only amplifies misinformation but also manifests in various forms, both online and offline~\cite{CASTANOPULGARIN2021101608,Johnson2019}
Such toxicity often takes the form of personal attacks, insults, and hateful rhetoric, exacerbated by the anonymity afforded by the online world emboldens users to shed social inhibitions, fostering a sense of detachment from the consequences of their words~\cite{doi:10.1089/cyber.2022.0005,10.1145/3372923.3404792}. Additionally, the lack of nonverbal cues, such as facial expressions or tone of voice, makes it easy to misinterpret communication, leading to misunderstandings and further escalations~\cite{Hess2023}. Civil discourse has become a casualty, being replaced by a hostile environment where respectful disagreement is a rarity~\cite{doi:10.1177/20563051231180638,10.1111/jcc4.12155}

Often prevalent in online interactions or social media platforms, language toxicity can contribute to a hostile environment, amplifying polarization, and undermining respectful discourse. Toxic speech can inflict harm even in the absence of slurs or epithets, with mechanisms that vary depending on the context and the target, creating a negative impact on the individuals involved \cite{tirrell2017toxic}. Additionally, the relatively low barriers to participation and the anonymity of online platforms can exacerbate this language toxicity, as individuals are more likely to express harmful views without fear of accountability \cite{gervais2023}. It can have significant psychological and social consequences, particularly when such language is directed toward individuals or groups based on their identity, beliefs, or behaviors \cite{volkow2021choosing}. For instance, the stigma created by harmful language—especially in contexts like mental health or addiction—can significantly worsen psychological distress and hinder social integration \cite{volkow2021choosing}. As polarization intensifies, the prevalence of toxic language on online platforms has led to growing pessimism regarding the future of public discourse~\cite{baum2020troll}. 
The proliferation of toxic language in digital spaces naturally breeds and amplifies collective pessimism ~\cite{chen2017online}. 

Online pessimism, particularly when expressed through toxic language, thrives in digital spaces where anonymity and distance obscure accountability. This pessimism feeds on negativity, cultivating a sense of disillusionment, frustration, and hopelessness surrounding societal, political, or personal issues. Toxic language exacerbates these emotions, drowning out constructive discourse and fostering an environment where vitriol reigns supreme~\cite{Coe2014}. Platforms like Twitter and Reddit, where emotional tones run high, illustrate this dynamic, with pessimism standing out as an emotion that is deeply influenced by social-psychological pressure~\cite{kumar-etal-2017-pessimist}.
Unlike individuals with positive dispositions, those with pessimistic tendencies exhibit negative cognitive processing of uncertain information~\cite{ISAACOWITZ2001255}, develop more negative thoughts that reduce their ability to handle stress and adjust mentally~\cite{Chang2002}, and demonstrate adverse online behavior~\cite{Wu2022}.
This online manifestation of pessimism aligns with broader psychological patterns in individual behavior.

While these toxic interactions and pessimistic expressions are often viewed through the lens of political divisions, our study reveals that polarization manifests across multiple dimensions in online discourse.
Beyond partisan differences, we observe polarization in user engagement (degree patterns), platform dynamics (between different social media architectures)~\cite{vandijck2013social}, linguistic expression (in language toxicity and sentiment), information complexity (through entropy variations)~\cite{Pilgrim2024}, and replying behaviors~\cite{Miyazaki2022}.
Understanding these various dimensions of polarization is crucial for comprehending how online discussions develop and fragment, extending beyond traditional political divides to encompass structural, behavioral, and communicative aspects of social media interaction.

Previous research has extensively explored the concepts of language toxicity and pessimism, but 
little research has focused on their role in the polarization of debunking efforts within social media environments. Existing literature has established that misinformation spreads rapidly on social media~\cite{DelVicario2016,Vosoughi2018,Lazer2018}, and that social media exhibit distinct patterns of engagement while debunking misinformation~\cite{Zhang2020,Cinelli2021}. However, it remains unclear how debunking contents vary in terms of language toxicity and sentiment, and how these variations influence broader social polarization. Understanding this dynamic is crucial for gaining insights into the development of hate speech, negative sentiment, misinformation, fake news, and the formation of echo chambers.
To address this gap, we investigated the polarized patterns of language toxicity and pessimism by analyzing discussions of the 2016 and 2020 U.S. presidential elections, and the QAnon conspiracy theory on Twitter and Reddit. 
This study sheds light on how toxicity and pessimism manifest within partisan communities, providing a more comprehensive understanding of the underlying dynamics.

Building on these observations of complex polarization patterns, the central objectives of this research are twofold.
First, we aim to map the polarized patterns of language toxicity and sentiment both within and between partisan groups—specifically Democrats and Republicans—on platforms like Twitter and Reddit. Second, we seek to explore how these linguistic features are linked to user engagement and behavior, with a particular emphasis on understanding the relationship between language toxicity and pessimism in debunking discourse.
 To guide our investigation, we pose the following research questions (RQs):
\textbf{RQ1} How do less-engaged users contribute to language toxicity and sentiment? \textbf{RQ2}  Whether the messaging mechanism of Twitter and Reddit contribute to the polarization? \textbf{RQ3}  How are replying behaviors correlated with language toxicity and pessimism?
By addressing the polarized patterns of language toxicity and sentiment in debunking discourse, this study offers valuable contributions to the ongoing conversation about the role of social media in shaping political and social polarization. Our findings have broad implications for both academic researchers and policymakers concerned with mitigating the harmful effects of toxic online behavior and fostering healthier, more constructive digital discourse.

\section*{Methods}
\subsection*{Data collection}
To investigate these research questions empirically, we collected data from two major social media platforms.
We used Twarc, a command line tool for collecting X data via the X API, to collect social data for our study. Over a 78-month period between October 1, 2016 and March 31, 2023, we used the Twitter Search API to gather tweets (now called ``posts'') by querying debunking-related keywords: ``fact check,'' ``fact-checking,'' ``fact checker,'' ``fact checkers,'' ``fake news,'' ``misinformation,'' ``disinformation,'' ``debunkers,'' ``debunker,'' ``debunking,'' and ``debunk.''. This dataset was used to extract tweets for the 2016, 2020 U.S. presidential elections and QAnon.
For Reddit, we used data maintained by Pushshift\endnote{\url{https://the-eye.eu/redarcs/}} from June 2005 to March 2023. The Pushshift Reddit dataset consists of two sets of files: submissions and comments~\cite{2001.08435}. Each line in a submission file contains a submission in JSON object format. The comments are a collection of NDJSON files, with each file containing one month of data. Each line in a comment file corresponds to a comment in JSON object format.



Our analysis covered Twitter and Reddit data spanning three significant topics. On Twitter, we examined three time windows from October 2016 to February 2017 (for the 2016 U.S. presidential election), August 2020 to January 2021 (for the 2020 U.S. presidential election), and April 2020 to May 2021 (for QAnon). The Twitter debunking dataset encompassed 86.7 million tweets in total: 7.1 million tweets related to the 2016 election, 23.1 million concerning the 2020 election, and 56.6 million addressing the QAnon.
For Reddit, we focused on similar time windows for the three topics. Our Reddit analysis revealed 461,318 debunking comments for the 2016 U.S. 
presidential election, 2,007,065 for the 2020 U.S. presidential election, and 2,237,749 for QAnon conspiracy theory, totaling 4.7 million debunking comments across all three periods.
By using topic-related keywords, we further obtained 2,084,881, 6,608,464, and 143,363 posts on Twitter, and 407,264, 157,917, and 14,637 comments on Reddit for the 2016, 2020 U.S. presidential elections and QAnon, respectively (cf. Table~\ref{S-tab:data} for details).

\subsection*{Sentiment calculation and pessimism detection}
We utilized NLTK\endnote{\url{https://www.nltk.org/}}, a well-recognized Python library for NLP, to calculate sentiment. The core idea for this calculation relies on the Valence Aware Dictionary and Sentiment Reasoner (VADER)~\cite{Hutto_Gilbert_2014}. The algorithm splits the text into individual words and assigns a score to each word to determine whether it is positive, negative, or neutral. Based on these word-level scores, VADER computes an overall sentiment score for the entire text. For sentiment analysis, we analyzed the aggregated text of each user. Specifically, for reply-to users, we aggregated the texts that reply-to users received. The sentiment, quantified by the $compound$ score, ranges from -1 (extremely negative) to 1 (extremely positive), indicating the overall polarity of the text.
For pessimism user identification, we employed a RoBERTa-based model~\cite{1907.11692}, specifically the ``cardiffnlp/twitter-roberta-base-emotion-multilabel-latest''\endnote{\url{https://huggingface.co/cardiffnlp/twitter-roberta-base-emotion-multilabel-latest}} checkpoint available on Huggingface. This model is one of the most widely adopted deep learning-based models for sentiment classification. It is a fine-tuned RoBERTa~\cite{camacho-collados-etal-2022-tweetnlp} model trained with Affect Tweets~\cite{mohammad-etal-2018-semeval} for emotion detection.
To obtain the proper $compound$ score and classification of a user, we aggregate the user’s text and then remove the duplicated texts.

\subsection*{Language toxicity measurement}
To define language toxicity in our analysis, we leverage the Perspective API~\cite{perspectiveapi}, a leading tool for the automated detection of toxic language.
Perspective API defines 
 language toxicity as ``rude, disrespectful, or unreasonable comments likely to drive someone away from a discussion''~\cite{Schramowski2022,Avalle2024}.
The Perspective API\endnote{\url{https://perspectiveapi.com/}} uses a probability score to indicate how likely it is that a reader would perceive the comment provided in the request as containing a given attribute. The score is a value between 0 and 1, with a higher score indicating a greater probability that a reader would perceive the comment as containing the attribute. For example, if a comment receives a probability score of 0.7 for attribute $TOXICITY$, indicating that 7 out of 10 people would perceive that comment as toxic.
To obtain the language toxicity value of a user, we aggregate the user’s
text and then remove the duplicated text.
For the replying scenario, each replied-to user's received texts were considered.
However, it is important to point out that such automated tools, like Perspective API, have limitations and potential biases, and people have to treat them considerably \cite{gehman-etal-2020-realtoxicityprompts}.

\subsection*{Identification of 1-degree users}
To determine how 1-degree users contributed to sentiment and language toxicity, we identified these users in the discussions of the 2016 and 2020 U.S. presidential elections.
We constructed retweet network for Twitter, and reply network for Reddit for each topic. In the retweet network, each node represented a user and directed edges between nodes represented retweets, and each node represented a user and directed edges between nodes represented replies for Reddit.
Then, we used NetworkX~\cite{Hagberg2008} to construct a 2-core network. Finally, 1-degree users are defined as the set difference between the total user population and the set of 2-core users.

\subsection*{Identification of Republican and Democratic users and verification of user classes}

As previous research~\cite{SIP-2021-0055} showed, the retweet network of Twitter can be used for user classification in QAnon conspiracy theory scenarios. We adopted this method, expecting to identify a characteristic retweet network where Republican and Democratic users are segregated. We constructed retweet networks using datasets from the 2016 and 2020 U.S. presidential elections and QAnon discussions. We applied k-core decomposition (k = 2)~\cite{6137224}, where each node represented a user and directed edges between nodes represented retweets. As expected, this resulted in a retweet network with two major clusters for each topic. We determined which cluster corresponded to Republican and Democratic groups by manually examining high-indegree users (those who were frequently retweeted) in each cluster, analyzing their tweets and profile descriptions.
However, due to the nature of the reply-based mechanism, it was not possible to build a retweet network for the Reddit dataset. To complement the network-based classification approach, we employed \textit{PoliticalBiasBERT}~\cite{baly2020we} to classify users as Republican or Democratic. \textit{PoliticalBiasBERT} is a BERT-based model\endnote{\url{https://huggingface.co/bucketresearch/politicalBiasBERT}} fine-tuned with articles and their political ideology annotations from Allsides\endnote{\url{https://www.allsides.com/}}.

To confirm whether the classification of Republican and Democratic users was reliable enough, we conducted a manual verification as follows. We conducted the manual verification by dividing all users into two classes. 
Two coders participated in this task.
On Twitter, 20 Republican users and 20 Democratic users were randomly selected from each topic of 2016, 2020 U.S. presidential elections and QAnon.
As a result, 120 users were obtained. 
On Reddit, we aggregated the users of all the topics and then randomly selected 25 Republican users and 25 Democratic users,  for a total of 50 users. 
Then, we checked the consistency of their classifications by computing Cohen’s kappa.
For Twitter, the resulting Kappa values were $0.9264$, $0.9116$, and $0.6482$ for the 2016, 2020 U.S. presidential elections, and QAnon, respectively.
For Reddit, the resulting Kappa values was $0.7248$.
Note that Cohen’s kappa value is interpreted as follows: 0.0–0.2 for
slight agreement; 0.2–0.4 for fair agreement; 0.4–0.6 for moderate agreement; 0.6–0.8 for
substantial agreement; and 0.8–1.0 for near-perfect agreement~\cite{kappa}.
The Kappa values indicated substantial agreement for QAnon, and near-perfect agreement for the 2016, 2020 U.S. presidential elections of Twitter, and substantial agreement for Reddit. 
The manual verification certified our user classification result as statistically reliable.

\subsection*{Identification of minimal entropy interval}
The Shannon entropy~\cite{shannon} was used to calculate the text entropy for each user: 
\begin{equation}
\centering
H=-\sum p(x)\log p(x)
\end{equation}, where \textit{H} represents the Shannon entropy, $p(x)$ represents the word probability of the text of a user.
To quantify the platform-specific differences in entropy, we measured the entropy difference between Twitter and Reddit.
We proposed the Minimal Entropy Interval Identification algorithm (Algorithm~\ref{alg:1}). The algorithm aims to identify text information patterns in user engagement on Twitter and Reddit by finding the minimal entropy interval containing the majority of users. This methodology provides a quantitative approach to understanding user text information concentration through entropy distribution analysis.
Given a distribution of entropy values from user interactions, we seek to find the smallest possible interval $[a, b]$ such that:
1. The interval contains $50\%$ (or just over $50\%$ due to calculation reasons) of all users
2. The interval length $l = b - a$ is minimized
3. The granularity of measurement is fixed at $0.1$.

\begin{algorithm}[htp]
\begin{algorithmic}[1]
\caption{Minimal Entropy Interval Identification}\label{alg:1}

\Function{ComputeProp}{$data, lower, upper$}
    \State Count the number of elements in $data$ that lie strictly between $lower$ and $upper$\;
    \State Divide the count by the total number of elements in $data$\;
    \State \Return the proportion
\EndFunction

\Function{FindIntervals}{$data, interval\_len$}
    \State $max\_value \gets$ maximum value in $data$
    \State $results \gets$ an empty list
    \For{each $start$ from 0 to $(max\_value - interval\_len)$ in steps of 0.1}
        \State $proportion \gets \text{ComputeProp}(data, start, start + interval\_len)$
        \If{$proportion > 0.5$}
            \State Append a record to $results$ containing:
            \State \hspace{1em} $start$: starting point of the interval
            \State \hspace{1em} $end$: $start + interval\_len$
            \State \hspace{1em} $length$: interval\_len
            \State \hspace{1em} $proportion$: computed proportion
        \EndIf
    \EndFor
    \State \Return $results$
\EndFunction

\Function{FindMinimumInterval}{$data$}
    \State $max\_value \gets$ maximum value in $data$
    \State $output \gets$ an empty list
    \For{each $interval\_len$ from 0 to $max\_value$ in steps of 0.1}
        \State Add results of $\text{FindIntervals}(data, interval\_len)$ to $output$
        \If{$output$ contains at least one result}
            \State \textbf{break}
        \EndIf
    \EndFor
    \State \Return $output$
\EndFunction
\end{algorithmic}
\end{algorithm}
 Minimal Entropy Interval Identification methodology was developed to quantitatively characterize and compare user behavior patterns across digital platforms through the systematic analysis of entropy distributions. The algorithm implements a three-tiered computational approach to identify the minimal interval containing the majority ($>50\%$) of users within an entropy distribution. The primary computational framework consists of three interconnected functions operating at different analytical levels. The base function, \texttt{ComputeProp}, calculates the proportion of users within specified entropy bounds by determining the ratio of users within a given interval to the total user population. Building upon this, the \texttt{FindIntervals} function employs a sliding window analysis technique with a granularity of $0.1$ to systematically identify intervals exceeding the $50\%$ threshold criterion. This function iteratively evaluates potential intervals across the entire entropy range, recording qualifying intervals along with their associated metrics including start point, end point, length, and contained proportion. The highest-level function, \texttt{FindMinimumInterval}, orchestrates the overall search process by implementing an incremental expansion strategy, beginning with the minimum interval length of $0.1$ and systematically increasing until a solution is found.

The hierarchical computational approach ensures the identification of the smallest possible qualifying interval, thereby providing a quantitative metric for behavioral concentration. 
This analytical framework enables quantitative cross-platform comparisons through the interpretation of interval lengths: smaller intervals indicate more concentrated user behavior patterns (clustered entropy values), while larger intervals suggest more dispersed behavioral patterns.
The methodology provides a robust foundation for comparative analysis of user engagement patterns across different platforms and topics, offering insights into platform-specific behavioral dynamics.

\subsection*{Bubble plot for entropy visualization}
The entropy magnitude for each platform is derived by computing the median entropy values of Republican and Democratic users for the 2016 and 2020 U.S. presidential elections and QAnon discussions on Twitter and Reddit.
Bubble size is a visualization size indicating the amount of entropy, here, we set $q = 500$, and a larger $q$ produces a larger size for normalization transformation.
The bubble size is calculated using:
\begin{equation}
   Size_{Bubble} = q \times \frac{2^\textit{H}}{2^{\min_{p,t}\textit{H}}}
\end{equation}, where a visualization parameter $q$ is used to a generate distinctive bubble size indicating the amount of entropy (Here, we set $q = 500$, and a larger $q$ produces a larger size for normalization transformation.), $\min_{p,t}\textit{H}$ represents the minimum entropy of a user category of a political affiliation for a topic ($t$) on a platform ($p$).
The difference of two entropies (${H_1}$, ${H_2}$) can be calculated using:

\begin{equation}
   \Delta H = H_1 - H_2 = \lg{\frac{Bubble_{Size_1}}{Bubble_{Size_2}}}
\end{equation}

The bubble size is explained as follows: If a text contains 256 distinct words, we obtain 8 bits for the text ($\lg256 = 8$), while we obtain 7 bits for 128 distinct words ($\lg128 = 7$), and so on.
In our bubble plot, this entropy-to-bubble-size transformation establishes a minimum bubble size of $q=500$ while preserving exponential relationships. For example, Reddit's Democratic users during the 2020 U.S. presidential election exhibited an entropy of $6.2$ (mapped to bubble size $1630$), compared to Twitter's Democratic users in 2016 with an entropy of $4.5$ (mapped to bubble size $500$). The bubble size reflects underlying differences in vocabulary size and platform-specific textual complexity.
These entropy are displayed as annotations on the visualization. 
Since this transformation is applied uniformly to all bubbles, it enables comparative analysis of relative entropy magnitudes across different platforms and topics.

\section*{Results}
\subsection*{Polarization in 1-Degree and 2-Core Users}
\begin{figure}[ht]
\centering
\includegraphics[width=0.9\linewidth]{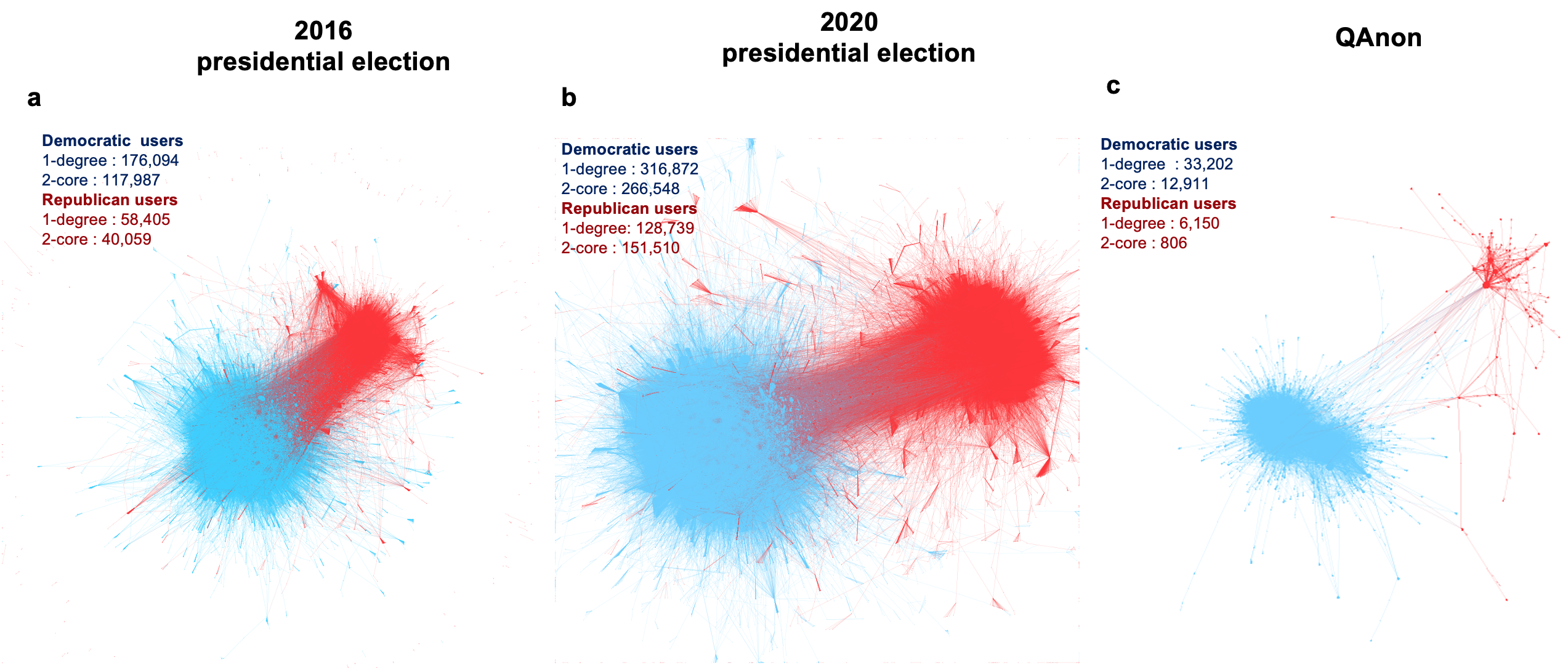}
\caption{Retweet network of 2016 (panel \textbf{a}), 2020 (panel \textbf{b}) presidential elections and QAnon (panel \textbf{c}) on Twitter.}
\label{fig:network}
\end{figure}

Figure~\ref{fig:network} shows the retweet network constructed from the
2016, 2020, U.S. presidential elections and QAnon dataset, revealing that Republican and Democratic users were segregated. 
Based on the retweet network analysis, we then identified and classified Republican and Democratic users on both Twitter and Reddit platforms (cf. Methods).
The demographics of the two classes of users were described in Table~\ref{tab:2core}.

We found that the majority of participants in discussions of the 2016, 2020 U.S. presidential elections, as well as QAnon, on both Twitter and Reddit, were 1-degree users. 
While 1-degree users constituted the numerical majority, our analysis revealed that 2-core users exhibited more pronounced polarization in both language toxicity and pessimism.
It is important to note that Twitter's degree is based on retweet networks, whereas Reddit's is derived from reply networks. 
As Table~\ref{tab:2core} shows, on Twitter, 340,234 (67.38\%), 566,813 (56.28\%), and 56,208 (80.68\%) of participants in discussions of the 2016, 2020 U.S. presidential elections and QAnon, respectively, were 1-degree users, compared to 164,728 (32.62\%), 440,393 (43.72\%), and 3,463 (19.32\%) 2-core users. A similar trend was observed on Reddit, where 61,376 (92.88\%), 164,505 (95.15\%), and 10,436 (99.24\%) of participants were 1-degree users, while 4,707 (7.12\%), 8,390 (4.85\%), and 80 (0.76\%) were 2-core users for the same topics, respectively.

\begin{table}[ht]
\centering
\caption{Demographics of 2-core and 1-degree users of Twitter and Reddit}
\begin{tabular}{cccc}
\hline
Platform                 & Topics                             & \#   2-core users (\%) & \#   1-degree users (\%) \\ \hline
\multirow{3}{*}{Twitter} & 2016 U.S. presidential election & 164,728   (32.62)      & 340,234   (67.38)        \\ \cline{2-4} 
                         & 2020 U.S. presidential election  & 440,393   (43.72)      & 566,813   (56.28)        \\ \cline{2-4} 
                         & QAnon                              & 13,463   (19.32)       & 56,208   (80.68)         \\ \hline
\multirow{3}{*}{Reddit}  & 2016 U.S. presidential election & 4,707   (7.12)         & 61,376   (92.88)         \\ \cline{2-4} 
                         & 2020 U.S. presidential election  & 8,390   (4.85)         & 164,505   (95.15)        \\ \cline{2-4} 
                         & QAnon                              & 80   (0.76)            & 10,436   (99.24)         \\ \hline
\end{tabular}
\label{tab:2core}
\end{table}

\begin{figure}[ht]
\centering
\includegraphics[width=0.8\linewidth]{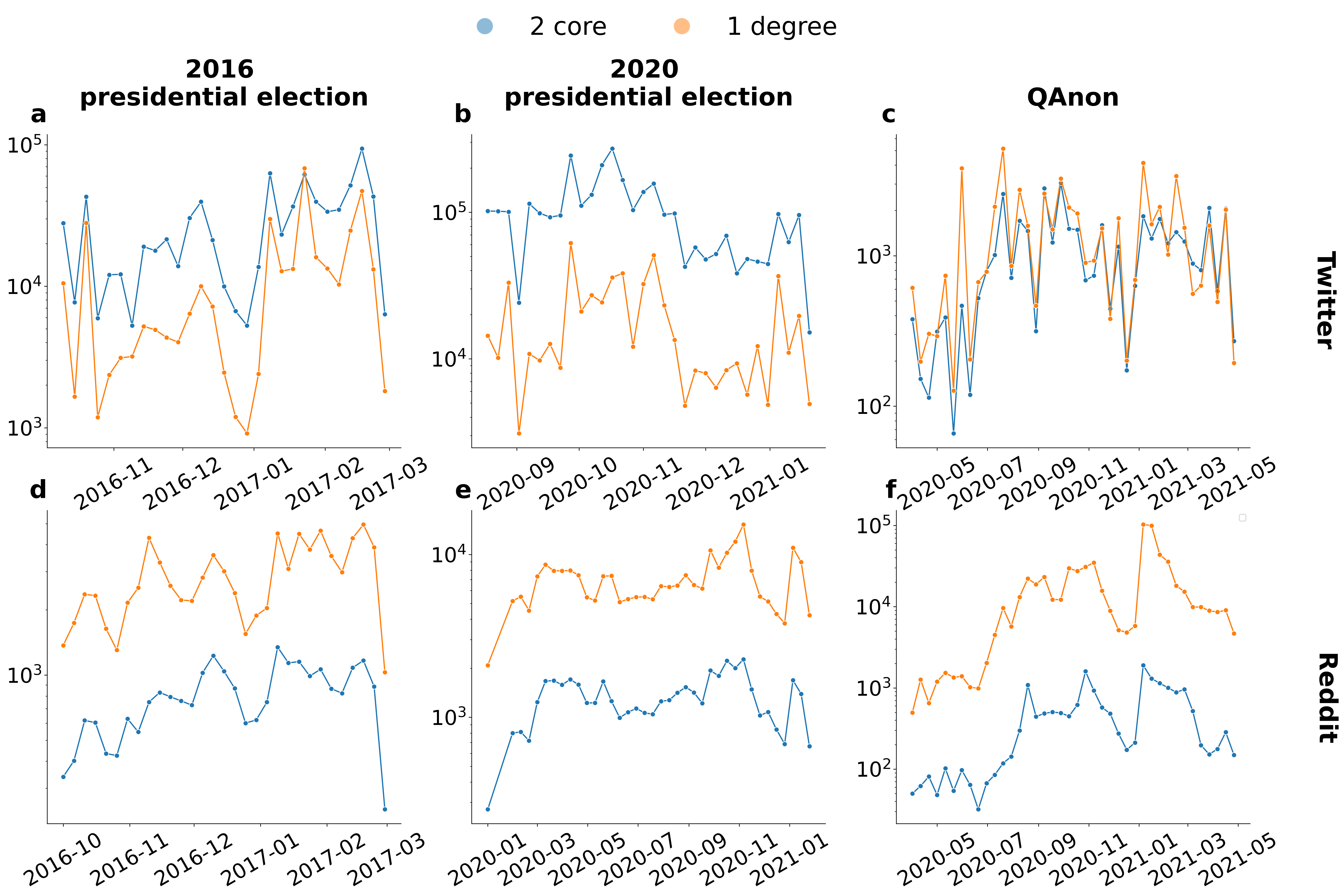}
\caption{Temporal dynamics of \#user patterns of 2-core and 1-degree user populations on Twitter and Reddit.
}
\label{fig:2core_temporal}
\end{figure}

Figure~\ref{fig:2core_temporal} illustrates the temporal oscillation across platforms and topics: for Twitter's 2016 and 2020 U.S. presidential election discussions and Reddit's 2016 U.S. presidential election topic, a 5-day segmentation was selected, while a 10-day segmentation was applied for Twitter's QAnon topic, the 2020 U.S. election topic on Reddit, and Reddit's QAnon topic. This approach keeps the analysis windows around 35 days across platforms and topics, allowing for a consistent comparison of trends. The oscillation shows that the patterns of 2-core and 1-degree users fluctuate following similar trends.
To statistically confirm this observation, we measured the Pearson correlation coefficient of temporal oscillations of 2-core and 1-degree users. The results are summarized in Table~\ref{S-tab:2core_pearson}. This reveals that 2-core and 1-degree users correlated with each other across topics. 
Further statistical analysis shows that, except QAnon, 2-core users are significantly more frequent than their 1-degree counterparts on Twitter ($\textit{P} < 0.01$ by Mann–Whitney U-test with a Bonferroni correction.), whereas 1-degree users are significantly more frequent than 2-core users across topics ($\textit{P} < 0.01$ by Mann–Whitney U-test with a Bonferroni correction.), suggesting a qualitative polarization in user engagement (Table~\ref{S-tab:2core_counts}).
To further investigate this difference between 2-core and 1-degree users, we examined whether similar significant differences are present in sentiment and language toxicity.

\begin{figure}[ht]
\centering
\includegraphics[width=\linewidth]{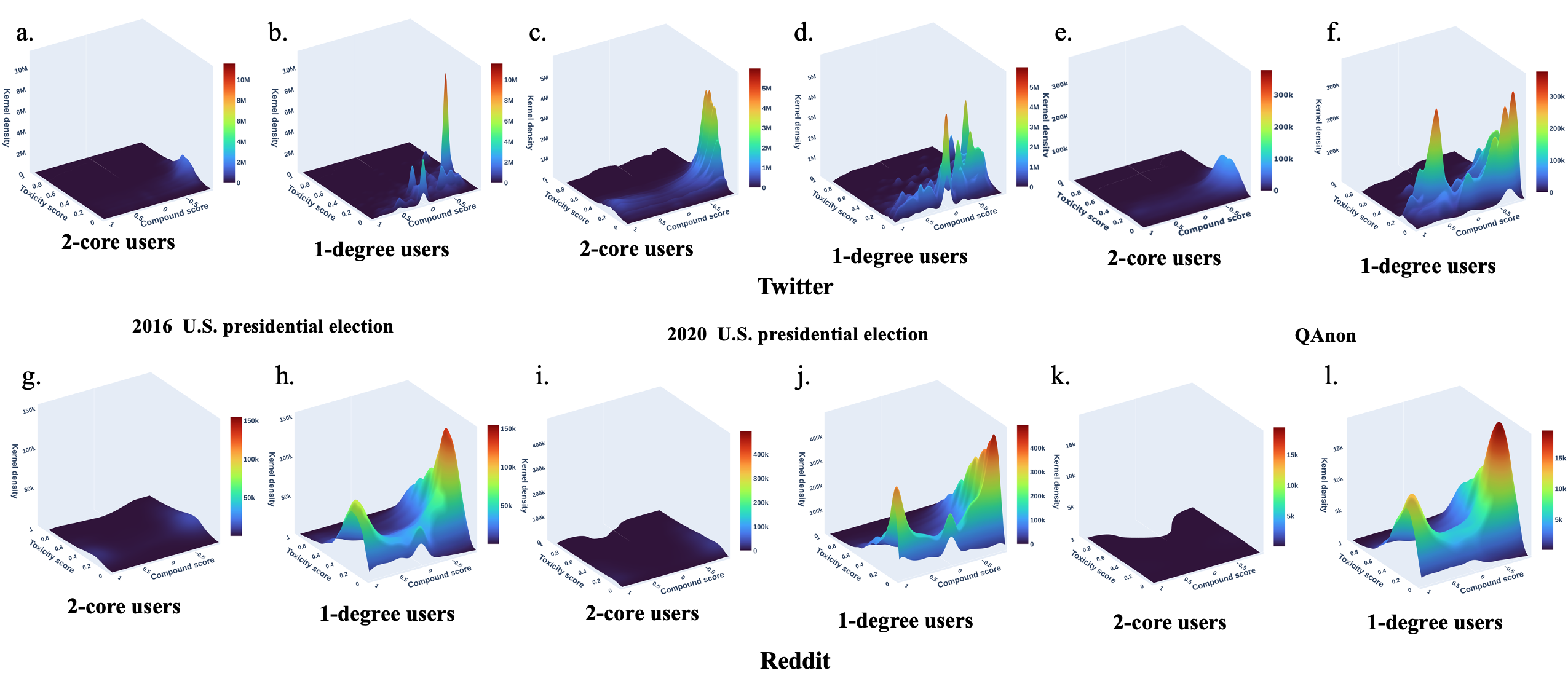}
\caption{Polarization of sentiment and language toxicity for 2-core users and 1-degree users across Twitter and Reddit platforms during the 2016, 2020 U.S. presidential elections, and QAnon topics. The 3D Gaussian distribution plots show the relationship between language toxicity (horizontal axis, $0$ is in the middle of the axis), $compound$ score (depth axis), and kernel density (vertical axis). The color gradient from dark blue to red represents the intensity of user engagement, with higher density values shown in warmer colors. This visualization compares behavioral patterns between 2-core users (more engaged users) and 1-degree users (peripheral users) across Twitter and Reddit platforms, revealing the distribution and concentration of polarized sentiments and language toxicity in debunking discourse. Higher kernel density values indicate a greater concentration of users exhibiting particular combinations of language toxicity and sentiment scores.}
\label{fig:2core}
\end{figure}

Since 1-degree users are the primary contributors to engagement through retweets and replies, we explored how their language patterns (language toxicity and sentiment) align with their engagement (degree status). To examine this, we analyzed the distribution of sentiment and language toxicity between user groups (2-core and 1-degree users) on both platforms (Figure~\ref{fig:2core}).
On Twitter, both 2-core and 1-degree users tend to cluster in areas of higher $compound$ scores with lower language toxicity, as shown in the bottom right of Figure~\ref{fig:2core}\textbf{a-f}. For Reddit, however, 2-core users displayed no significant language patterns, while 1-degree users exhibited notable patterns in regions of lower language toxicity (Figure~\ref{fig:2core}\textbf{g-l}. Among these Reddit 1-degree users, the sentiment was polarized: for users with a language toxicity level between 0.2 and 0.4, 30\% demonstrated positive sentiment ($0.5 <= compound <= 1$), whereas 60\% exhibited negative sentiment ($-1 <= compound <= -0.5$). This indicates a clear emotional polarization among low-language-toxicity 1-degree users on Reddit.

Further validating these observations, Hartigan’s Dip Test confirmed a statistically significant multimodal distribution of sentiment scores among 1-degree users engaged in discussions of the 2016, 2020 U.S. presidential elections and the QAnon topic on Twitter and Reddit ($\textit{P} < 0.01$ by Mann–Whitney U-test with a Bonferroni correction.) (Table~\ref{S-tab:1}). This statistical analysis result reinforces the argument that the distribution of sentiment and language toxicity among 1-degree users is indeed polarized, supporting the hypothesis of polarized sentiment landscapes within these discussions.

In alignment with these statistical findings, a deeper examination of the joint distribution of language toxicity and sentiment scores for 1-degree users reveals a consistent trend of emotional polarization across both platforms. As demonstrated in Figures~\ref{fig:2core}\textbf{b, d, f} for Twitter and Figures~\ref{fig:2core}\textbf{h, j, l} for Reddit, the distribution exhibits a clear bimodal pattern (Table~\ref{S-tab:bimodal}) across the topics. For instance, on Twitter, in the 2016 U.S. presidential election (Figure~\ref{fig:2core}\textbf{b}), $15.53\%$ of 1-degree users with a language toxicity range of $0.04--0.4$ express positive sentiment ($0 <= compound <= 0.5$), while $40.91\% $show negative sentiment ($-0.75 <= compound <= -0.4$)

with a substantial 56.44\% of users falling into the bimodal category.
In the 2020 U.S. election discussion (Figure~\ref{fig:2core}\textbf{d}), the pattern continues with 31.87\% exhibiting positive sentiment and 33.94\% showing negative sentiment, bringing the bimodal category to 65.81\%. For the QAnon topic (Figure~\ref{fig:2core}\textbf{f}), a similar bimodal trend appears, where 20.76\% of users display positive sentiment and 36.27\% display negative sentiment, resulting in 57.03\% of bimodal users among 1-degree participants.
On Reddit, this polarization is similarly pronounced. For the 2016 U.S. presidential election (Figure~\ref{fig:2core}\textbf{h}), 21.48\% of 1-degree users show positive sentiment, while 34.26\% exhibit negative sentiment, with bimodal users comprising 55.74\%. Similar patterns are observed for the 2020 U.S. election and QAnon topics, with Reddit’s 1-degree users consistently showing more than 50\% bimodal distribution, highlighting the substantial emotional polarization across these discussions and further validating the bimodal nature of sentiment among these users.

Having identified the sentiment patterns, we next examined the association between pessimism and language toxicity to better understand emotional dynamics.
We first found that there was a substantial difference in the distribution of language toxicity and pessimism between 2-core and 1-degree users across different topics using the Fligner-Killeen Test on Twitter and Reddit (except the pessimism for QAnon on Reddit).
This difference is quantified using Cliff’s delta to measure the effect size of language toxicity between 1-degree and 2-core users on each platform (Table~\ref{S-tab:effect_size}).
The effect sizes suggest a medium to large impact of user degree levels on language toxicity, with substantial polarization between 1-degree and 2-core users in topics such as the 2016 U.S. presidential election on Reddit ($\textit{Cliff's Delta} = -0.960$), the 2020 U.S. presidential election on Reddit ($\textit{Cliff's Delta} = -0.714$), the 2020 U.S. presidential election on Twitter ($\textit{Cliff's Delta} = -0.972$), and the QAnon topic on Twitter ($\textit{Cliff's Delta} = -0.966$).
The effect sizes demonstrate significant differences in language toxicity between Democratic and Republican users, particularly in the context of U.S. presidential election discussions. These partisan differences are identified on both Reddit and Twitter. For example, during the 2016 U.S. presidential election discussions on Reddit, the effect size suggests a marked difference in language toxicity levels between Democratic and Republican users ($\textit{Cliff's Delta} = -0.781$). This pattern continues in the 2020 U.S. presidential election, with a substantial effect size observed on Reddit ($\textit{Cliff's Delta} = -0.661$) and an even greater difference on Twitter ($\textit{Cliff's Delta} = -0.862$). These results indicate that political leaning significantly influences language toxicity, with strong polarization observed across platforms.
Unlike language toxicity, the effect sizes suggest political leaning has a smaller impact on pessimism (Table~\ref{S-tab:effect_size2}).

\begin{figure}[ht]
\centering
\includegraphics[width=0.8\linewidth]{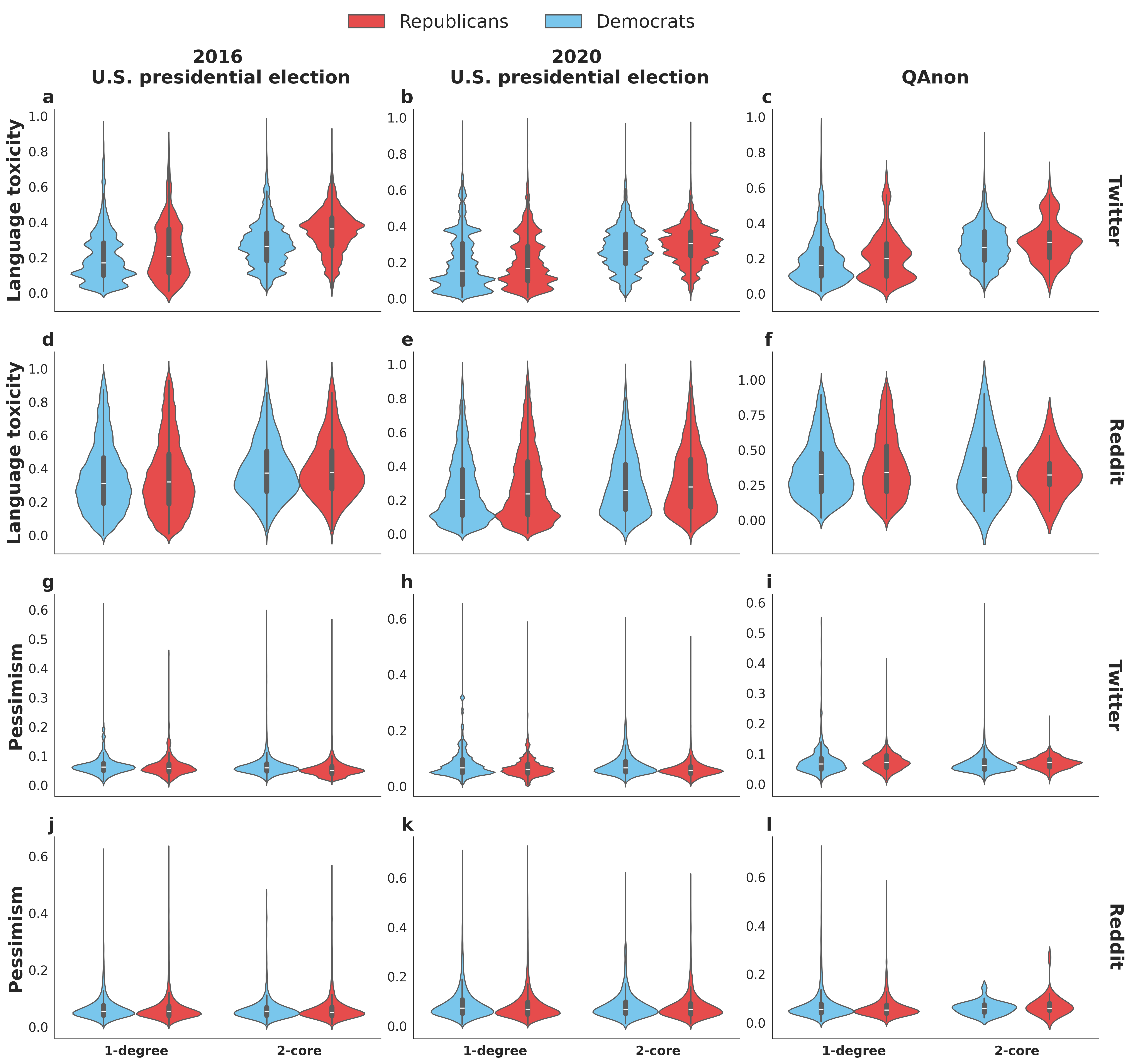}
\caption{
The difference in toxicity and pessimism between Republican and Democratic 1-degree and 2-core users across the 2016, 2020 U.S. presidential elections and QAnon on Twitter and Reddit. Panels (\textbf{a-f}) show toxicity distributions: Twitter (panel \textbf{a-c}) and Reddit (panel \textbf{d-f}), while panels (\textbf{g-l}) show pessimism distributions: Twitter (panel \textbf{g-i}) and Reddit (panel \textbf{j-l}). The width of each violin indicates the density of observations at that value, with the black bars showing the interquartile range and the white dot representing the median.
}
\label{fig:2core_1degree}
\end{figure}

To explore the difference in terms of degree, we examined language toxicity (Figure~\ref{fig:2core_1degree} \textbf{a-f}), pessimism (Figure~\ref{fig:2core_1degree} \textbf{g-i}) and entropy (Figure~\ref{fig:2core_1degree} \textbf{m-r}) levels from the perspective of political leaning.
We found that, except for 2-core users of the QAnon topic, the Republican users show significantly higher language toxicity  ($\textit{P} < 0.01$, Table~\ref{S-tab:user_type}) and pessimism ($\textit{P} < 0.01$, Table~\ref{S-tab:user_type2}) than Democratic users. 
Within the 1-degree and 2-core user categories, the majority of Democratic users exhibited significantly higher pessimism levels than Republican users ($\textit{P} < 0.05$, except for 2-core Reddit users in the 2016 U.S. presidential election, QAnon users on Twitter, and QAnon users in the 2-core category on Reddit; Table~\ref{S-tab:user_type2}).
These results suggest that Republican and 2-core users are more prone to engage in toxic discourse, while Democratic and 1-degree users are more prone to engage in pessimistic discourse.

\subsection*{Negative relationship between language toxicity and pessimism} 

\begin{figure}[htp]
\centering
\includegraphics[width=1\linewidth]{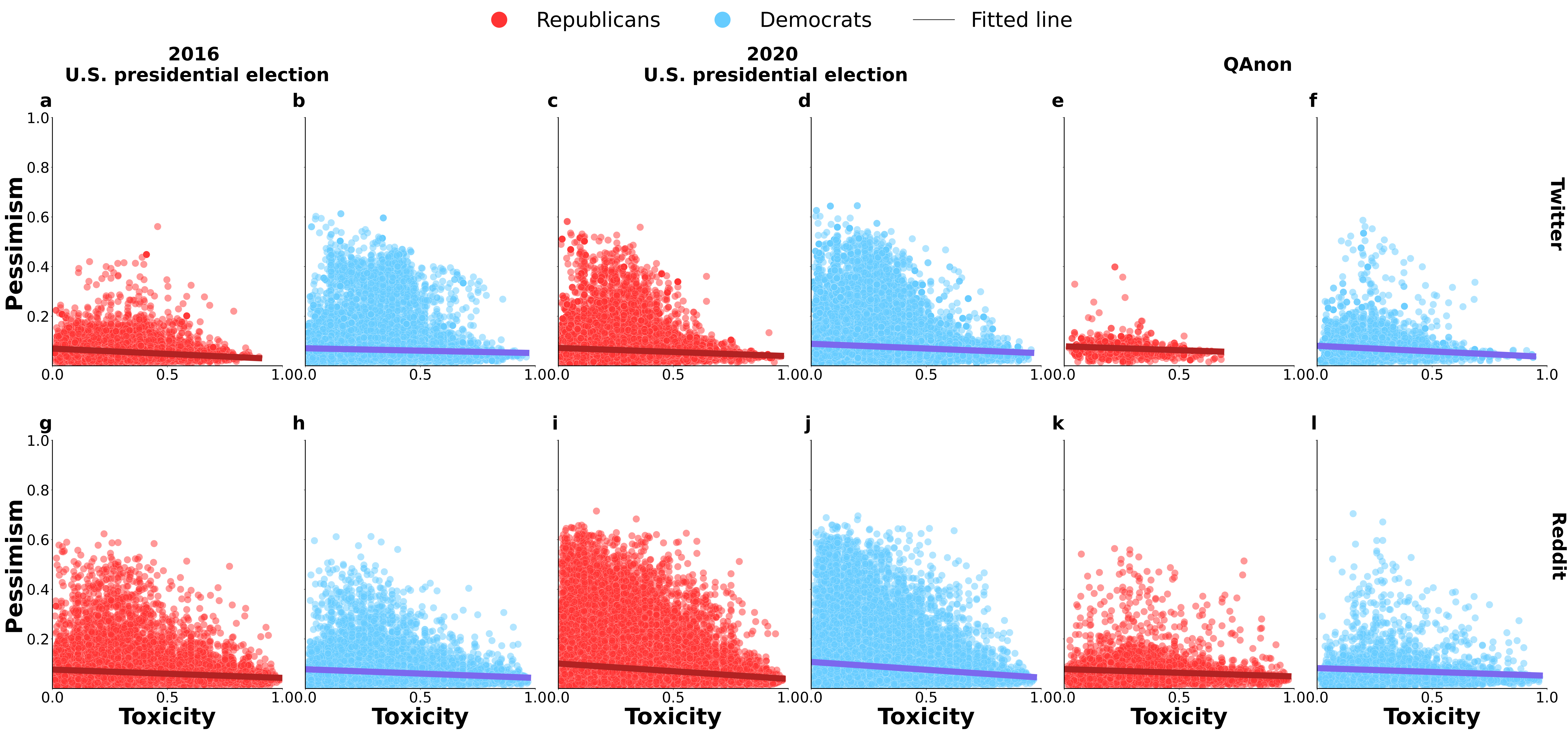}
\caption{Toxicity as a function of pessimism for Republican (red) and Democratic (blue) users across different political events on Twitter (panel \textbf{a-f})) and Reddit (panel \textbf{g-l}). For all panels, fitted lines represent the best-fit linear regression lines; Error bands represent 95\% confidence intervals. The plots reveal a consistent negative correlation between toxicity and pessimism across both platforms and political affiliations. 
}
\label{fig:pessimism_toxicity}
\end{figure}

Contrary to our initial expectations of a positive correlation between toxicity and pessimism, we discovered that users exhibiting higher language toxicity levels actually tend to express lower levels of pessimism across topics for both Democratic and Republican users on Twitter and Reddit (Figure~\ref{fig:pessimism_toxicity}).

Our analysis of the relationship between language toxicity and pessimism across topics and platforms reveals a consistent negative correlation. Higher levels of toxicity are associated with lower levels of pessimism.
This trend is consistent across discussions on Twitter about the 2016 U.S. presidential election (Figure~\ref{fig:pessimism_toxicity}\textbf{a, b}), the 2020 U.S. presidential election (Figure~\ref{fig:pessimism_toxicity}\textbf{c, d}), and QAnon (Figure~\ref{fig:pessimism_toxicity}\textbf{e, f}). Similarly, on Reddit, this trend appears in discussions of the 2016 U.S. presidential election (Figure~\ref{fig:pessimism_toxicity}\textbf{g, h}), the 2020 U.S. presidential election (Figure~\ref{fig:pessimism_toxicity}\textbf{i, j}), and QAnon (Figure~\ref{fig:pessimism_toxicity}\textbf{k, l}).
Interestingly, the distribution of pessimism remains clustered at lower levels across all toxicity values, indicating that users express relatively low pessimism, even when exhibiting a wide range of toxic language.

To statistically confirm the negative relationship between toxicity and pessimism, we analyzed Pearson correlation coefficients (Table~\ref{S-tab:pessimism_toxicity}).
On Twitter, Republican users consistently show a stronger negative relationship between language toxicity and pessimism than Democratic users across all topics ($\textit{P} < 0.001$). For the 2016 U.S. presidential election, the correlation for Republicans is stronger ($r = -0.216$) than for Democrats ($r = -0.079$). Similarly, in the 2020 U.S. presidential election, Republicans display a stronger negative correlation ($r = -0.137$) compared to Democrats ($r = -0.126$). This trend is even more evident in QAnon discussions, where the correlation for Republicans is ($r = -0.312$) and for Democrats is ($r = -0.155$), suggesting that, for Republicans on Twitter, higher language toxicity is more strongly linked to lower pessimism. On Reddit, the trends show slight variation. In the 2016 U.S. presidential election, Democratic users exhibit a slightly stronger negative relationship ($r = -0.148$) than Republicans ($r = -0.144$). However, for the 2020 U.S. presidential election and QAnon discussions, Republicans display a stronger negative relationship, with correlations of ($r = -0.194$) and ($r = -0.113$), respectively, than Democrats at ($r = -0.169$) and ($r = -0.097$). Overall, these Pearson correlation coefficients indicate that Republican users on both platforms generally exhibit a stronger negative correlation between language toxicity and pessimism, suggesting that, for these users, increased language toxicity tends to be more consistently linked with reduced pessimism.

\begin{figure}[htp]
\centering
\includegraphics[width=0.9\linewidth]{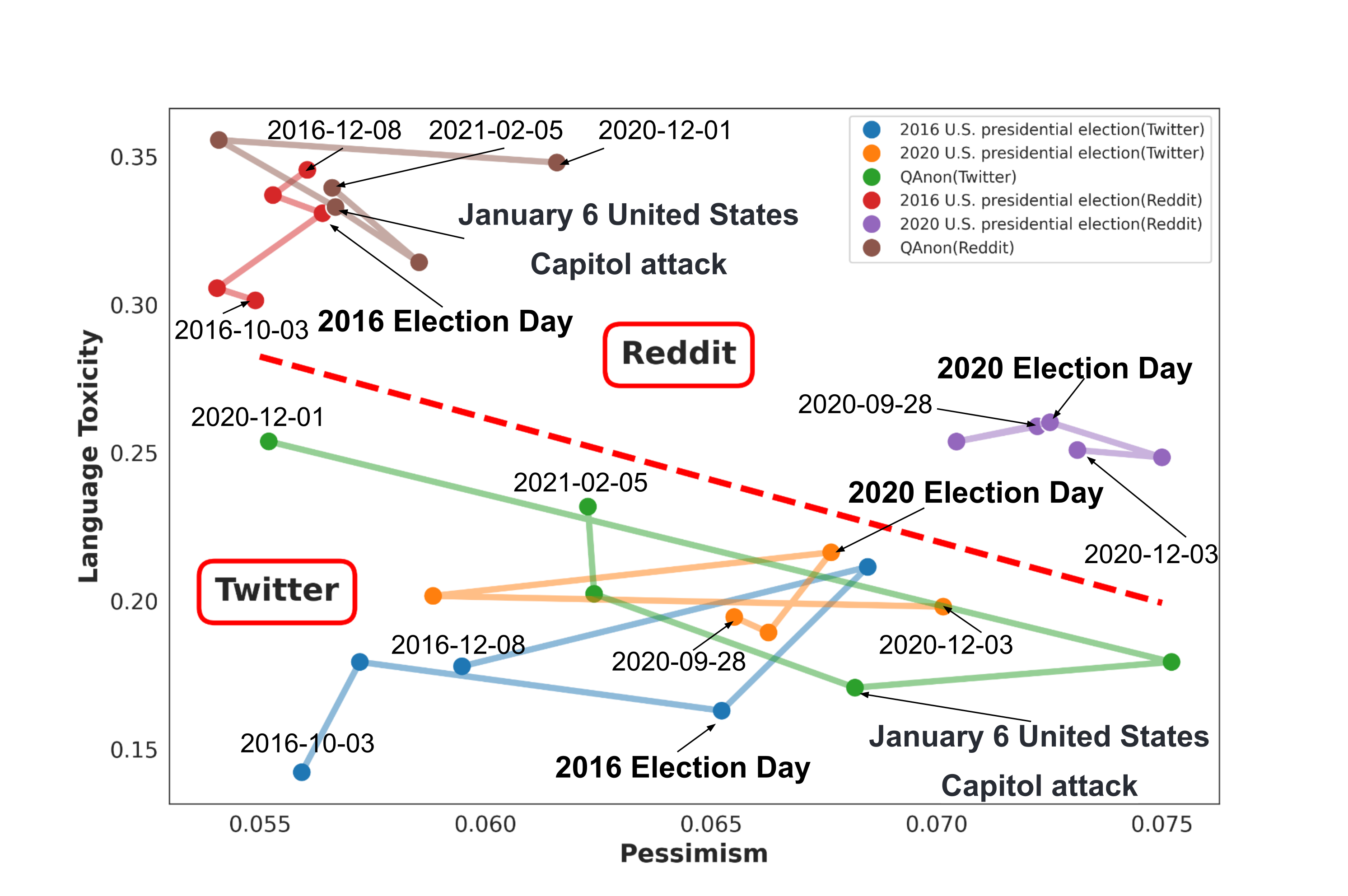}
\caption{The temporal polarization of language toxicity and pessimism across Twitter and Reddit. Our analysis involved several steps to extract and process retweet data, allowing us to examine trends in language toxicity and pessimism over time. First, we filtered timestamped retweet data from the full dataset, isolating tweets generated each day and grouping them by date, with all tweets from a given day consolidated into one group. For each daily group, we aggregated text data by user and calculated language toxicity, and pessimism (cf. Methods for details). To address outliers, we conducted a boxplot quartile analysis, filtering out extreme values before calculating the mean language toxicity and sentiment scores for each user, which resulted in a daily sequence of values in the format (Pessimism, Language toxicity, Date). Focusing on specific events such as 2016 Election Day, 2020 Election Day and the January 6 Capitol attack, we extracted data within a 15-day window surrounding each event date (from 7 days before to 7 days after). Each date in this period, such as 2020-03-12, represents the median score computed over a 15-day window centered on that date. The red dashed line is the regression line ($k=-4.158, b=0.511$) between language toxicity and pessimism.
This approach generated a time series that illustrates shifts in language toxicity and pessimism, with five key data points per topic to capture changes in sentiment and language toxicity as each event unfolded.
}
\label{fig:TimesliceLine}
\end{figure}

Inspired by Phillips (1958)~\cite{Phillips1958}, we used a scatter diagram (Figure~\ref{fig:TimesliceLine}) to illustrate the polarized spatio-temporal features of language toxicity and pessimism across topics on Twitter and Reddit (Figure~\ref{fig:TimesliceLine}).
The results reveal a clear pattern of platform polarization between Reddit and Twitter in terms of the relationship between pessimism and language toxicity. Specifically, there is an observable negative relationship between pessimism and language toxicity, with higher pessimism generally linked to lower language toxicity levels, as identified in Figure~\ref{fig:pessimism_toxicity}.
We found that language toxicity on Twitter were significantly lower than on Reddit ($\textit{P} < 0.001$, Mann-Whitney U Test).
By contrast, the pessimism of the 2016 and 2020 U.S. presidential elections on Twitter was significantly lower than on Reddit ($\textit{P} < 0.001$, Mann-Whitney U Test), but the pessimism of QAnon is significantly larger than Twitter's counterpart.

As shown in Figure~\ref{fig:TimesliceLine}, language toxicity and pessimism displayed distinct temporal patterns during major political events on both platforms.
During the 2016 U.S. presidential election, Twitter demonstrated relatively low levels of language toxicity ($\sim 0.15$) and pessimism ($\sim 0.057$), whereas Reddit showed substantially higher language toxicity ($\sim 0.30$) and pessimism ($\sim 0.065$), indicating a more contentious tone in its discussions. By the 2020 U.S. presidential election, language toxicity on Twitter increased to $\sim 0.25$, with pessimism remaining moderate at $\sim 0.060$. In contrast, Reddit maintained higher language toxicity ($\sim 0.30$) and slightly elevated pessimism ($\sim 0.067$). 
The January 6 Capitol attack triggered distinct patterns across social media platforms. Reddit experienced peak language toxicity (measured by hostile language) of $\sim0.33$ with low pessimism ($\sim0.056$), while Twitter showed lower toxicity ($\sim0.17$) but higher pessimism ($\sim0.068$). The higher toxicity on Reddit suggests more confrontational debates occurred there, while Twitter's pattern indicates immediate emotional reactions without escalating into hostile exchanges.
This suggests that Reddit exhibited stronger polarization during these events.

\subsection*{Polarization on Platforms}
The platform mechanisms of Twitter and Reddit differ significantly. Twitter, with its focus on real-time interactions and concise communication, tends to amplify polarizing content, while Reddit's structure of topic-specific subreddits supports more focused and sustained discussions. 

These differences lead us to assume that the messaging mechanism of the two platforms affects the polarization of political leaning, entropy of users, language toxicity and pessimism.
To focus on the majority of users with sufficient entropy. We proposed the Minimal Entropy Interval Identification method to examine the entropy expressed by 50\% (or just over 50\% due to calculation) of the users of a topic on a platform (Table~\ref{tab:entropy}; cf. Methods for The  Minimal Entropy Interval Identification method is described in Algorithm~\ref{alg:1}.)
For each topic, the minimal entropy interval of Twitter is smaller than that of Reddit, respectively.
The minimal entropy interval indicates that discussions on Twitter convey a more limited amount of information, reflecting a narrower and more uniform focus, whereas discussions on Reddit involve a broader entropy range, suggesting a higher level of informational diversity and complexity.

\begin{table}[ht]
\centering
\caption{Entropy minimal interval for 50\% of users}
\label{tab:entropy}
\begin{tabular}{llrrrr}
Platform & Topic & \multicolumn{1}{l}{Entropy start} & \multicolumn{1}{l}{Entropy end} & \multicolumn{1}{l}{Minimal entropy interval} & \multicolumn{1}{l}{User rate (\%)} \\ \hline
Twitter & 2016 U.S. presidential election & 4.1 & 4.8 & 0.7 & 51.3 \\
Twitter & 2020 U.S. presidential election & 4.5 & 5.7 & 1.2 & 50.1 \\
Twitter & QAnon                      & 4.8 & 5.4 & 0.6 & 53.4 \\
Reddit  & 2016 U.S. presidential election & 4.4 & 6.6 & 2.2 & 50.1 \\
Reddit  & 2020 U.S. presidential election & 5.3 & 6.9 & 1.6 & 50.2 \\
Reddit  & QAnon                      & 4.5 & 6.4 & 1.9 & 51.2
\end{tabular}
\end{table}

To statistically confirm the platform difference,
we measured the entropy of Republican and Democratic on Twitter and Reddit (Table~\ref{S-tab:figure8-entropy}). On Reddit, Democratic users consistently show higher entropy than Republican users($\textit{P} < 0.001$). This means Democratic users shared much more diverse posting patterns than Republicans. On Twitter, there is no significant difference between the two groups during election discussions. But for the QAnon topic, Democratic users again show much higher entropy than the Republican users ($\textit{P} < 0.001$). Overall, Democratic users tend to share more varied communication styles. Republican users show more uniform patterns. This partisan gap is wider on Reddit than on Twitter.
The observed patterns indicate that platform architecture plays a crucial role in shaping the manifestation of political entropy polarization.

\begin{figure}[htp]
\centering
\includegraphics[width=0.8\linewidth]{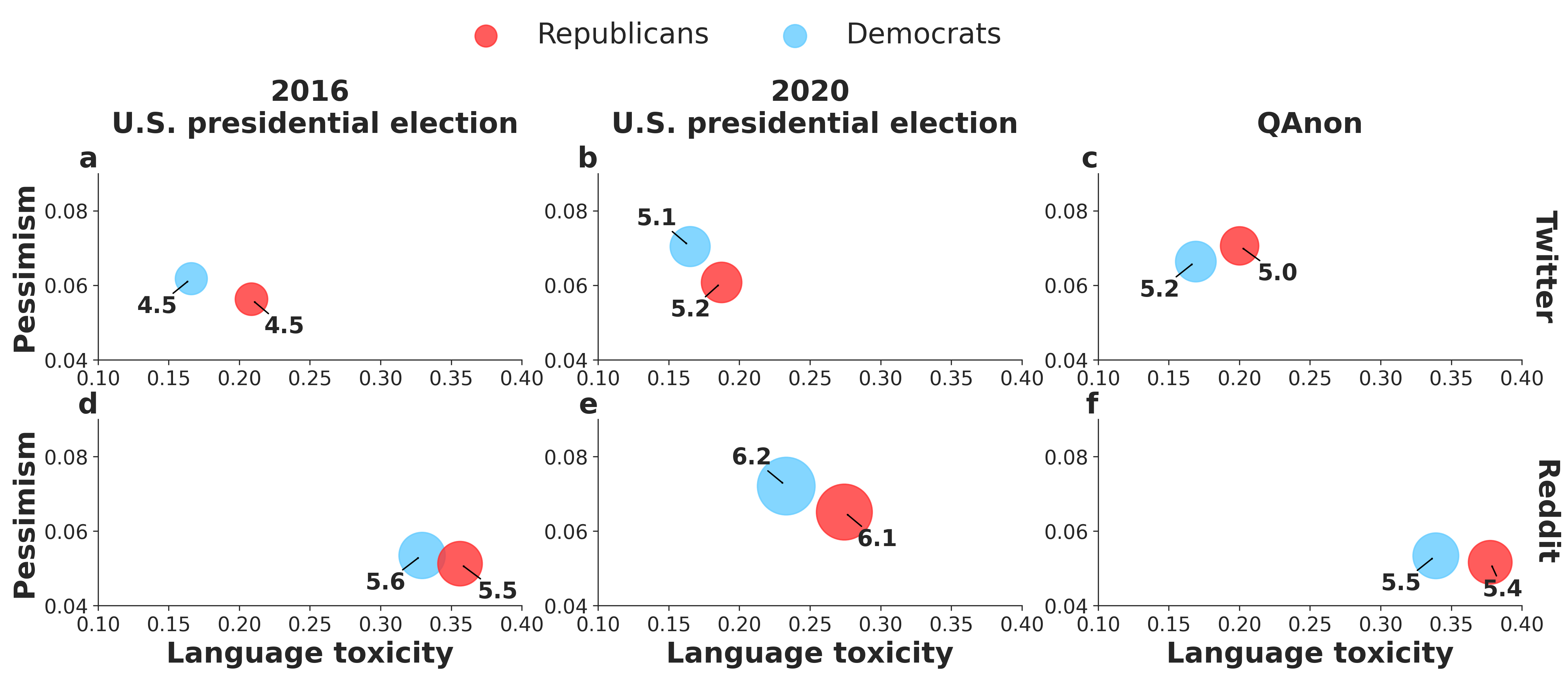}
\caption{Entropy median polarization analysis of users in the range of entropy minimal interval (cf. Methods for details) of Republican and Democratic. The figure is organized by topic (2016 presidential election, 2020 presidential election, and QAnon) and platform (Twitter, panel (\textbf{a–c}); Reddit, panel (\textbf{d-f})), with Republican in red and Democratic in blue. The bubble size indicates a normalized entropy value, with the entropy value annotated on each bubble. Each bubble size was determined by the entropy of each bubble (cf. Methods for details). The x-axis quantifies language toxicity, while the y-axis represents pessimism levels.}
\label{fig:entropy_polarization}
\end{figure}

To examine how platform polarization relates to entropy, language toxicity, and pessimism, we visualized entropy as bubble sizes in Figure~\ref{fig:entropy_polarization} (cf. Methods).
Remember that language toxicity negatively correlates with pessimism, suggesting a platform-based polarization (Figure~\ref{fig:TimesliceLine}).
Here, the median measurements indicate that Twitter shows lower language toxicity and lower entropy than Reddit.
For example, during the 2016 U.S. presidential election, on Twitter  (Figure~\ref{fig:entropy_polarization}\textbf{a}), Democratic users exhibited lower entropy ($4.5$) and language toxicity($\sim0.17$), while on Reddit  (Figure~\ref{fig:entropy_polarization}\textbf{d}), the counterparts showed higher entropy ($5.6$) and language toxicity ($\sim0.33$). The republican users showed a similar tendency. 
For the 2020 U.S. presidential election, the entropy medians are also lower on Twitter than on Reddit for both political affiliations (Figure~\ref{fig:entropy_polarization}\textbf{a,e}).
The polarization toxicity of QAnon was evident, with Twitter users on the far left (Figure~\ref{fig:entropy_polarization}\textbf{c}) while Reddit users are on the far right (Figure~\ref{fig:entropy_polarization}\textbf{f}).  
To statistically confirm the observations, we measured the differences using the Kolmogorov-Smirnov test. The results are summarized in Table~\ref{S-tab:kstest}. The entropy, language toxicity, and pessimism were significantly different between Twitter and Reddit ($\textit{P} < 0.001$).

\begin{figure}[htp]
\centering
\includegraphics[width=0.9\linewidth]{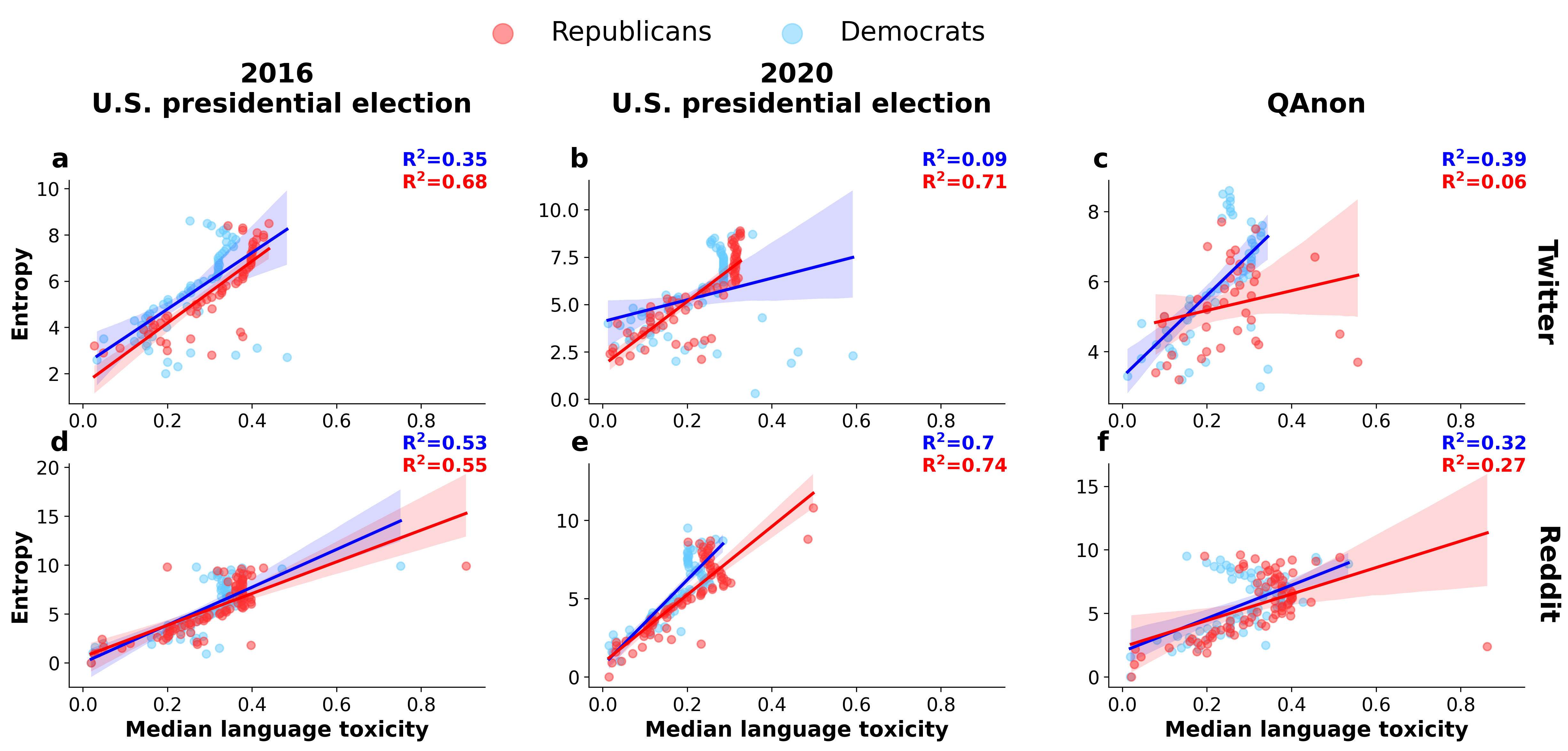}
\caption{Entropy-median language toxicity polarization. The figure is categorized by topic (2016 U.S. presidential election, 2020 U.S. presidential election, and QAnon) and platform (Twitter, panel \textbf{a–c} and Reddit, panel \textbf{d-f}). The data is split by political affiliation, with Republicans represented by red and Democrats by blue. For each panel, $R^2$  values indicate the strength of correlation between entropy and toxicity levels.The shaded areas represent 95\% confidence intervals for the regression lines. The increasing slopes across all panels indicate a positive relationship between entropy and language toxicity, though this relationship varies in strength across platforms and political affiliations.
}
\label{fig:entropy_toxicity}
\end{figure}

The entropy and language toxicity polarization of Twitter and Reddit inspired us to investigate their association with political affiliation.
We found that language toxicity and entropy showed a positive correlation (Figure~\ref{fig:entropy_toxicity}). Reddit consistently exhibited higher median language toxicity levels, reaching up to $\sim$0.8 compared to Twitter's maximum of $\sim$0.6, and showed stronger correlations between entropy and language toxicity as indicated by higher $R^2$ values.
For political leaning, Twitter demonstrated more pronounced partisan differences in toxicity-entropy correlations. For the 2016 U.S. presidential election (Figure~\ref{fig:entropy_toxicity}\textbf{a}), Republican users ($R^2 = 0.68$) and Democratic users ($R^2 = 0.35$) showed a correlation difference of $0.68-0.35=0.33$. This partisan gap widened to $0.71-0.09=0.62$ for 2020 (Figure~\ref{fig:entropy_toxicity}\textbf{b}), and reached $0.39-0.06=0.33$ for QAnon discussions. Mann-Whitney U tests further supported these partisan differences, showing Democratic users on Twitter exhibited significantly lower language toxicity in both election discussions ($P = 0.0$, Table~\ref{S-tab:figure8-toxicity}).
In contrast, Reddit exhibited minimal partisan differences in these correlations, with gaps of $0.55-0.53=0.02$, $0.74-0.70=0.04$, and $0.27-0.32=-0.05$ for the 2016, 2020 presidential U.S. elections, and QAnon respectively. However, Democratic users on Reddit consistently showed higher entropy across all topics ($P < 0.05$, Table~\ref{S-tab:figure8-entropy}) and lower toxicity ($P < 0.05$, Table~\ref{S-tab:figure8-toxicity}). While Reddit's discussion structure appeared to facilitate higher overall toxicity levels, Twitter's messaging mechanism seemed to amplify partisan differences in how users engaged with complex discussions.

\subsection*{Polarization of replying}
\begin{figure}[ht]
\centering
\includegraphics[width=0.9\linewidth]{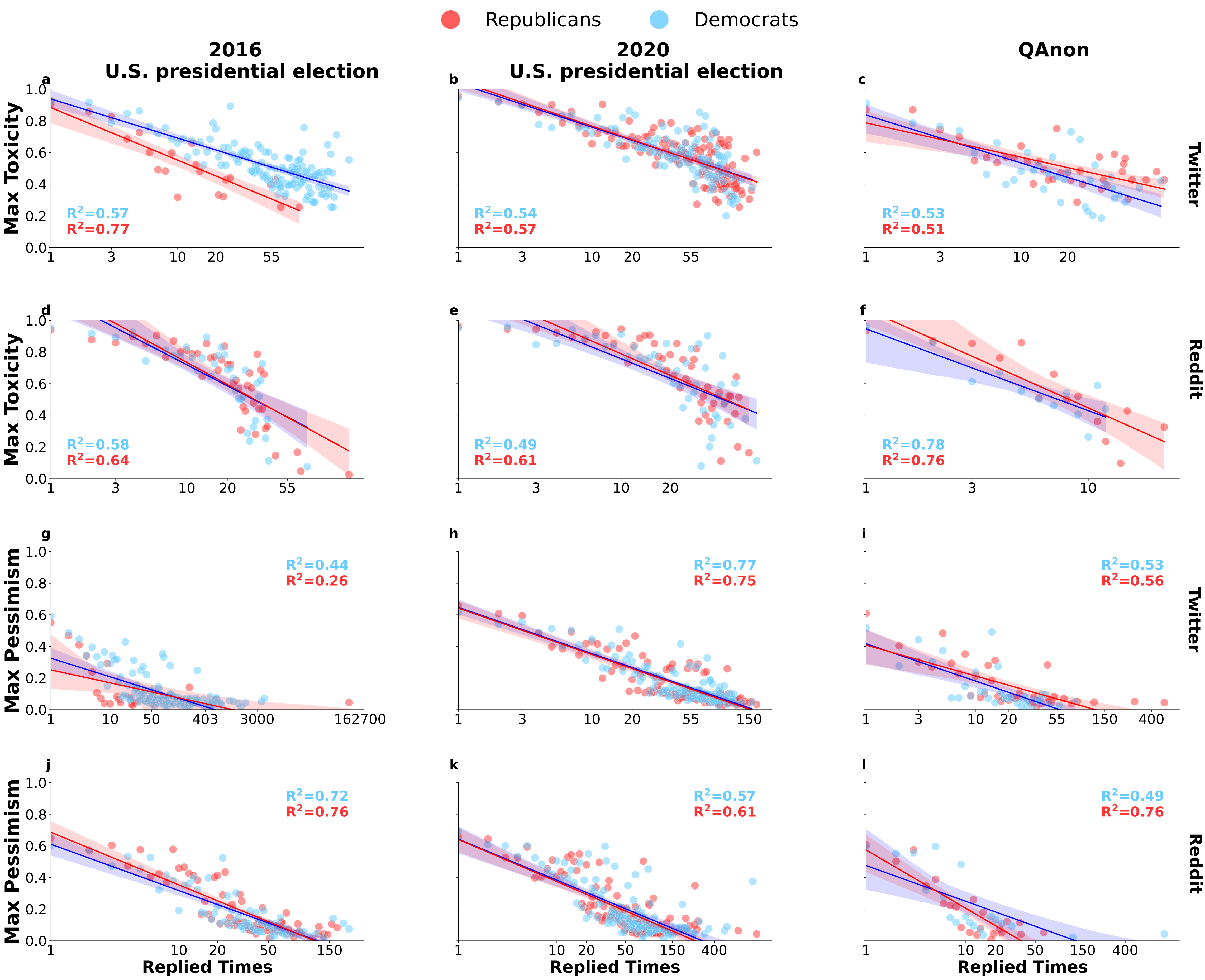}
\caption{The maximum toxicities and pessimism of replies of Republican and Democratic replied-to users received across different political events and platforms. The X-axis, replied times, is in log scale to better visualize the distribution across different engagement levels. The Y axis indicates the maximum language toxicity (panel \textbf{a-f}) and pessimism (panel \textbf{g-l}). Reds indicate Republican replied-to users, and blues indicate Democratic replied-to users. Shaded areas represent 95\% confidence intervals for the fitted regression lines. The data reveals that while both toxicity and pessimism generally decrease with more replies.
}
\label{fig:polarization_reply}
\end{figure}

Platform polarization is particularly evident when comparing Reddit and Twitter, with Reddit showing consistently steeper slopes in both language toxicity and pessimism reduction. We found that in the 2016 U.S. presidential election discussions on Twitter, both Democrats and Republicans show moderate toxicity reduction (Figure~\ref{fig:polarization_reply}\textbf{a}), which continues in the 2020 U.S. presidential election (Figure~\ref{fig:polarization_reply}\textbf{b}) and QAnon discussions (Figure~\ref{fig:polarization_reply}\textbf{c}). In contrast, Reddit exhibits stronger toxicity reductions across all topics, as shown in Figure~\ref{fig:polarization_reply}\textbf{d-f}. According to Table~\ref{S-tab:regression_toxicity}, in QAnon discussions on Reddit, Republicans show a steeper decline in toxicity (slope $a = -0.27$) compared to Twitter (slope $a = -0.09$). This is further supported by the $R^2$ values from Figure~\ref{fig:polarization_reply}\textbf{f} and Figure~\ref{fig:polarization_reply}\textbf{c}, with Reddit showing stronger correlations ($R^2 = 0.78$ for Republicans, $R^2 = 0.76$ for Democrats) compared to Twitter ($R^2 = 0.51$ for Republicans, $R^2 = 0.53$ for Democrats).

The steeper slopes and higher $R^2$ values observed on Reddit, compared to Twitter, indicate that Reddit's structured, community-driven interactions and moderation policies are more effective in reducing extreme content than Twitter's real-time, algorithm-driven dynamics.
Polarization based on political leaning shapes these trends distinctly across platforms. During the 2020 U.S. presidential election on Reddit, Table~\ref{S-tab:regression_toxicity} shows Republican and Democratic users have similar steep declines in toxicity ($a = -0.20$ for Republicans, $a = -0.18$ for Democrats) and Table~\ref{S-tab:regression_pessimism} indicates similar patterns in pessimism ($a = -0.12$ for Republicans, $a = -0.11$ for Democrats), but with different intercept ($b$ values). Republicans start with higher language toxicity ($b = 1.24$) than Democrats ($b = 1.17$).
As shown in Figure~\ref{fig:polarization_reply}\textbf{b} and Figure~\ref{fig:polarization_reply}\textbf{h}, on Twitter, the slopes are more moderate but nearly identical between parties for both toxicity ($a = -0.12$ for both) and pessimism ($a = -0.13$ for both), suggesting that platform characteristics may influence partisan behavior more than political affiliation alone.

Analysis of individual topics revealed that the effectiveness of polarization mitigation varied significantly across different political contexts.
QAnon discussions show the most pronounced differences between platforms, with Reddit exhibiting the steepest toxicity reduction slopes ($a = -0.27$ for Republicans, $a = -0.22$ for Democrats) in Table~\ref{S-tab:regression_toxicity} compared to Twitter's more modest declines ($a = -0.09$ for Republicans, $a = -0.13$ for Democrats), as visualized in Figure~\ref{fig:polarization_reply}\textbf{c} and Figure~\ref{fig:polarization_reply}\textbf{f}. The 2016 and 2020 U.S. presidential election topics show more consistent patterns across platforms, with the 20120 U.S. presidential election displaying similar toxicity reduction slopes on Twitter ($a = -0.12$ for both parties, Figure~\ref{fig:polarization_reply}\textbf{b}) and slightly steeper slopes on Reddit ($a = -0.20$ for Republicans, $a = -0.18$ for Democrats, Figure~\ref{fig:polarization_reply}\textbf{e}). The pessimism metrics in Table~\ref{S-tab:regression_pessimism} follow similar patterns but with generally smaller slopes, particularly in the 2016 U.S. presidential election discussions on Twitter ($a = -0.04$ for Republicans, $a = -0.05$ for Democrats, Figure~\ref{fig:polarization_reply}\textbf{g}).
The consistently smaller slopes for pessimism ($a = -0.04$ to $-0.05$ on Twitter, compared to toxicity slopes of $a = -0.12$) indicate that pessimistic content is more resistant to reply-based moderation than toxic content.

Previous research showed that Democratic users received more language toxicity than Republican users on Twitter~\cite{xu2024}. Our cross-platform analysis reveals that this pattern varies by platform. According to Table~\ref{S-tab:replying_toxicity}, on Twitter, Democratic users indeed received significantly more toxic responses in both the 2016 and 2020 U.S. presidential elections ($\textit{P} < 0.05$). However, Reddit shows an inverse pattern where Democratic users consistently received significantly less toxic replies across all topics ($\textit{P} < 0.05$). Meanwhile, they showed less pessimistic responses on Twitter particularly in the 2016 U.S. presidential election discussions ($\textit{P} < 0.001$) as shown in Table~\ref{S-tab:replying_pessimism}, but experienced higher levels of pessimism on Reddit ($\textit{P} < 0.05$). The QAnon discussions present a unique case: while toxicity differences become non-significant on Twitter ($p \approx 1.0$), pessimism differences remain significant ($\textit{P} < 0.001$).
These statistical findings support our regression analysis and suggest that platform messaging mechanisms might shape user interaction patterns and emotional expression.

\section*{Discussion} 
Our longitudinal analysis of social media debunking efforts reveals important insights into the manifestation of language toxicity and pessimism. Although our study focused specifically on Twitter and Reddit discussions of the 2016 and 2020 U.S. presidential elections and QAnon conspiracy theory, prior research on polarization patterns across social media platforms \cite{Vosoughi2018} suggests broader applicability of our findings.
This broader applicability is supported by research showing consistent patterns of polarization across different social media environments and various controversial topics \cite{Vosoughi2018}.
Building on these observations, our analysis demonstrates that polarization in debunking discourse operates through multiple interconnected mechanisms, extending beyond traditional partisan divisions to encompass user engagement patterns, platform dynamics, and emotional expression.

Regarding \textbf{RQ1} about the role of less engaged users
In addressing \textbf{RQ1} regarding how less engaged users contribute to language toxicity, our analysis revealed that peripheral participants (1-degree users) play a previously underappreciated role in shaping the polarized landscape of debunking discussions.
While research has traditionally focused on echo chambers and highly engaged users as primary vectors for spreading misinformation, our findings indicate that peripheral participants play a crucial role in shaping language toxicity of debunking discourse. This phenomenon can be understood through the lens of deindividuation theory from social psychology \cite{lowry2016adults}: when individuals feel less accountable due to limited community investment, they may express more extreme views when attempting to correct misinformation \cite{reicher1995social}. However, our binary classification of user engagement (1-degree versus 2-core) may oversimplify the spectrum of participation patterns in debunking activities, as users might shift between categories over time or display different behaviors across topics.

Our \textbf{RQ2} examined whether the messaging mechanisms of Twitter and Reddit contribute to polarization. We found that platform architecture significantly influences polarization patterns, with Reddit's discussion structure facilitating higher overall toxicity levels while Twitter's messaging mechanism amplifies partisan differences. This aligns with research showing how platform features shape information exposure \cite{bakshy2015exposure}. The negative correlation between language toxicity and pessimism was particularly pronounced on Reddit, suggesting that its community-driven structure may intensify emotional expressions in debunking efforts. These platform-specific patterns may generalize to emerging social media platforms with similar architectural features.

In examining \textbf{RQ3}, we found a significant negative correlation between reply frequency and toxicity levels, with this effect being particularly pronounced on Reddit.
This suggests that sustained interaction in debunking efforts might activate social calibration mechanisms \cite{doi:10.1177/0735275112437163}, similar to how intergroup contact reduces prejudice in offline settings \cite{Bail2018}. Yet Yarchi et al. (2021) \cite{yarchi2021political} found that increased interaction doesn't always reduce polarization in fact-checking contexts, highlighting the complexity of these dynamics.

Building on Brady et al.'s (2017) ~\cite{brady2017emotion} framework of emotional content's role in political information diffusion, our study reveals three critical mechanisms that drive polarization in debunking discourse: the outsized influence of peripheral users in generating toxic debunking content, the platform-dependent relationship between information complexity and polarized debunking discourse, and the distinct patterns of emotional expression across partisan lines when challenging misinformation. 
Although our analysis centered on U.S. presidential elections and the QAnon conspiracy theory across Twitter and Reddit, the fundamental and platform-independent nature of these three mechanisms suggests broader applicability: they reflect basic patterns of human behavior in online information sharing rather than platform-specific phenomena
The consistent patterns we observed suggest these findings could be applicable to emerging platforms like TikTok, Instagram, or YouTube, where similar dynamics of user engagement and information sharing exist. Additionally, the mechanisms we identified - particularly regarding user engagement levels and emotional expression - likely manifest in other controversial topics such as climate change \cite{Cook2019}, abolitionism~\cite{10.1093/actrade/9780190213220.001.0001}, and \#Gamergate~\cite{gamegate}.
Based on these three mechanisms, we propose that effective intervention strategies should address: (1) early engagement of peripheral users to prevent toxic content generation, (2) platform-specific architectural adjustments to manage information complexity, and (3) partisan-aware moderation approaches that account for different styles of emotional expression.
Platform designers and policymakers should consider these dynamics when developing strategies to reduce polarization in debunking efforts, particularly by creating features that accommodate different partisan styles of emotional expression in debunking discussions.

\section*{Acknowledgements}
This study is partially supported by the 2023-2024 Annual Seed Research Funding of the Department of Science and Technology of Communication, University of Science and Technology of China (\#ASRF2023). We also thank Prof. Zhiwen Hu of Zhejiang Gongshang University for the fruitful discussion.

\section*{Author contributions statement}
Wentao Xu conceptualized the research and designed the experiments. Wenlu Fan, Tenghao Li, and Shiqian Lu performed the experimental work, implemented the code, and created the visualizations. Bin Wang facilitated data collection. Wentao Xu analyzed the results and prepared the initial manuscript draft. All authors reviewed and approved the final manuscript.

\setcounter{endnote}{1}  
\renewcommand{\notesname}{Notes}  
\theendnotes  




\bibliography{main}

\end{document}